\documentstyle{article}
\newcommand{\be}{\begin{eqnarray}}
\newcommand{\ee}{\end{eqnarray}}
\newcommand{\nn}{\nonumber \\ \nonumber \\}
\newcommand{\nl}{\\  \nonumber \\}

\renewcommand{\vec}[1]{\mbox{\boldmath$#1$}}

\begin{document}

\pagenumbering{arabic}

\title{Spin-1/2 Particles in Non-Inertial Reference Frames:  Low- and
High-Energy Approximations}
\author{D. Singh$^1$, G. Papini$^2$ \\
$^1$Department of Physics, University of Alberta, \\
Edmonton, Alberta, T6H 2J1, Canada  \\
$^2$Department of Physics, University of Regina, \\
Regina, Saskatchewan, S4S 0A2, Canada}
\date{}

\maketitle

\begin{abstract}
Spin-1/2 particles can be used to study inertial and gravitational effects
by means of interferometers, particle accelerators, and ultimately
quantum systems.  These studies require, in general, knowledge of the
Hamiltonian and of the inertial and gravitational quantum phases.  The
procedure followed gives both in the low- and high-energy
approximations.  The latter affords a more consistent treatment of mass
at high energies.  The procedure is based on general relativity and on a
solution of the Dirac equation that is exact to first-order in the metric
deviation.  Several previously known acceleration and rotation induced
effects are re-derived in a comprehensive, unified way.  Several new
effects involve spin, electromagnetic and inertial/gravitational fields in
different combinations.
\end{abstract} 

\setcounter{section}{0}
\setcounter{equation}{0}


\section{Introduction}

The behaviour of quantum systems in inertial and gravitational fields
is of interest in investigations regarding the structure of spacetime
at the quantum level \cite{***}. Quantum objects are in fact finer and
more appropriate probes of structures that appear classically as results of
limiting procedures. Though a definitive answer to questions regarding
the
fundamental structure of spacetime may only come from a successful
quantum theory of gravity, the extrapolation of general relativity from
planetary lengths, over which it is well established, to Planck's
length requires a leap of faith in its validity of over forty orders of
magnitude and the resolution of difficult quantization problems. The
alternative, performing experiments at Planck's length, appears remote
indeed. A more modest, but realistic approach consists in verifying the
theory at intermediate lengths. This may be accomplished, to some
extent, by considering the interaction of classical inertial and
gravitational fields with quantum objects. A vast array of effects can
be predicted in this instance and a unified treatment is afforded by
Einstein's theory. General relativity incorporates the equivalence
principle from the outset and observations, where feasible, do 
confirm that inertia and gravity interact with quantum systems in ways
that are compatible with Einstein's views. This is borne out of
measurements on superconducting electrons \cite{1} and on neutrons
\cite{3} which are
certainly not tests of general relativity per se, but offer 
tangible evidence that the effect of inertia and Newtonian gravity on
wave functions down to lengths of $10^{-3}$ and $10^{-13}$ cm
respectively is that predicted by wave equations compliant with 
general relativity \cite{DW,2}.

Inertial effects must be identified with great accuracy. This is
dictated by their unavoidable presence in Earth-bound and near-space
experiments of ever increasing accuracy aimed at testing fundamental
theories. They also
provide a guide in the study of relativity because, in all instances where 
non-locality is not an issue, the
equivalence
principle, in some of its forms \cite{LA,MA}, ensures the existence of a 
gravitational effect for each
inertial effect.  

Among the quantum mechanical probes, spin-1/2 particles play a
prominent role and in reality some of the most precise experiments in
physics involve Dirac particles. They are very versatile tools that can
be used in a variety of experimental situations and energy ranges while
still retaining essentially a non-classical behaviour. Within the context 
of general relativity  comprehensive studies of the Dirac equation 
were conducted by De Oliveira and Tiomno \cite{DE} and Peres \cite{14a}. 
More recently,
spin-inertia and spin-gravity interactions \cite{4} have been shown to 
have
non-trivial physical and astrophysical consequences.
This is the case for neutrinos whose inertia/gravity interactions are just
starting to be studied \cite{5}.  Superconducting and neutron
interferometers of
large dimensions \cite{6,7} hold
great promise in many of these investigations. They can provide accurate
measurements of quantum phases, whose role is important in gyroscopy,
and possibly in testing general relativity \cite{8,9}.
It is anticipated that similar studies  will be performed with particle
accelerators \cite{10}.  The forerunner of this second group of
investigations is
the work of Bell and Leinaas \cite{11} in which evidence was found for
the
coupling of spin to rotation.

For most studies involving non-relativistic Dirac particles, 
the inertial/gravitational phase is of paramount importance.  For other
problems, such as the interaction of inertia/gravitation at the atomic
level \cite{!}, 
the knowledge of the Hamiltonian is of greater importance.  The
derivation of the Hamiltonian is usually accomplished by following a
sequence of Foldy-Wouthuysen (FW) transformations \cite{DE}.  It has been
recently presented in comprehensive form by Hehl
and Ni \cite{12} purely within the framework of special relativity and in
the local
frame of the fermion.  In the present work, the non-relativistic case is
tackled by means of a procedure that renders the quantum phase
manifest and can, therefore, be applied to both types of problems. 

For tests involving accelerators, a high-energy approximation
corresponding to the FW-transformation was given long ago by Cini and
Touschek (CT) \cite{CT} for free Dirac particles. Their work is
extended here to
include external electromagnetic and gravitational fields and quantum
phases to first order.  The derivation of the Hamiltonian can then be
accomplished in a standard way.

The method and derivation of the Hamiltonian are given in Section 2.
The low-energy approximation in Section 3 uses the FW-transformation
and follows the standard textbook approach \cite{13}.  In contrast, the
high-energy approximation in Section 4 uses the CT-transformation and
involves a non-standard procedure.


\section{Derivation of the Dirac Hamiltonian}

We use the formalism of general relativity, that encompasses both
inertial and true gravitational fields, to derive the Hamiltonian of a
fermion in a non-inertial frame.  

The starting point is represented by the covariant Dirac equation
\cite{14}
\be
\left(i \gamma^\mu (x) D_\mu -
\frac{mc}{\hbar}\right)\psi (x) & = & 0,
\label{2.1}
\ee
where the generalized matrices $\gamma^{\mu} (x)$ satisfy the relation
$\left\{\gamma^\mu (x), \gamma^\nu (x) \right\} \ = \ 2g^{\mu \nu}
(x)$ and are related to the usual Dirac matrices
$\gamma^{\hat{\alpha}}$ by
means of the vierbeins $e^\mu{}_{\hat{\alpha}} (x)$.  Caratted indices
refer to the observer's
local inertial frame.  In (\ref{2.1}), $D_\mu \ \equiv \ \nabla_\mu + i
\Gamma_\mu, \ \psi (x)$ is the wavefunction 
defined for a general co-ordinate frame,
$\nabla_\mu$ represents the usual 
covariant derivative, $\Gamma_\mu$ is the spinor connection which follows
from $D_\mu \gamma_\nu (x) \ = \ 0$ and is given
by 
\be
\Gamma_\mu  \ = \ \frac{i}{4} \gamma^\nu (\nabla_\mu \gamma_\nu) 
\ = \ - \frac{1}{4} \sigma^{\hat{\alpha} \hat{\beta}}
e^\nu{}_{\hat{\alpha}}
 (\nabla_ \mu e_{\nu \hat{\beta}}),
\label{2.1a}
\ee
where $ \sigma^{\hat{\alpha} \hat{\beta}} \ = \ \frac{i}{2} [
\gamma^{\hat{\alpha}},
\gamma^{\hat{\beta}} ].$  Obviously, $\nabla_\mu \psi(x) \ = \ \partial_\mu
\psi(x),$ where $\partial_\mu$ indicates partial differentiation.

It is possible to define a local co-ordinate frame according to an
orthonormal tetrad with three-acceleration $\vec{a}$ along a particle's
world-line and three-rotation $\vec{\omega}$ 
of the
spatial triad, subject to Fermi-Walker transport.  This tetrad
$\vec{e}_{\hat{\mu}}$, is related to the general co-ordinate tetrad 
$\vec{e}_\mu$ by
\be
\vec{e}_{\hat{0}} & = &  \left(1 + \frac{\vec{a} \cdot \vec{x}}{c^2}
\right)^{-1} \left[ \vec{e}_{0} - \frac{1}{c} (\vec{\omega} \times
\vec{x})^k \vec{e}_k \right],
\label{2.3}
\nl
\vec{e}_{\hat{\imath}} & = & \vec{e}_i .
\label{2.4}
\ee
The corresponding vierbeins relating the two frames are then
\be
e^0{}_{\hat{0}} & = &  \left(1 + \frac{\vec{a} \cdot \vec{x}}{c^2}
\right)^{-1},
\label{2.5}
\nl
e^k{}_{\hat{0}} & = &  - \frac{1}{c} \left(1 + \frac{\vec{a} \cdot
\vec{x}}{c^2} \right)^{-1} \epsilon^{ijk} \, \omega_i \, x_j,
\label{2.6}
\nl
e^0{}_{\hat{\imath}} & = & 0,
\label{2.7}
\nl
e^k{}_{\hat{\imath}} & = & \delta^k{}_i .
\label{2.8}
\ee
Similarly, by inverting (\ref{2.3}) and (\ref{2.4}), we find the inverse
vierbeins
\be
e^{\hat{0}}{}_{0} & = &  \left(1 + \frac{\vec{a} \cdot \vec{x}}{c^2}
\right)
\label{2.9}
\nl
e^{\hat{k}}{}_{0} & = &  \frac{1}{c} \epsilon^{ijk} \, \omega_i \, x_j,
\label{2.10}
\nl
e^{\hat{0}}{}_{i} & = & 0,
\label{2.11}
\nl 
e^{\hat{k}}{}_{i} & = & \delta^k{}_i.
\label{2.12}
\ee
The vierbeins satisfy the orthonormality conditions 
\be
\delta^{\hat{\alpha}}{}_{\hat{\mu}} & = &  e^\nu{}_{\hat{\mu}}
e^{\hat{\alpha}}{}_{\nu},
\label{2.13}
\nl
\delta^\alpha{}_{\mu} & = &  e^{\hat{\nu}}{}_{\mu}
e^\alpha{}_{\hat{\nu}}.
\label{2.14}
\ee
It follows that the metric tensor components are
\be
g_{00} & = & \left(1 + \frac{\vec{a} \cdot \vec{x}}{c^2} \right)^2 
+ \frac{2}{c^2}\left[ \left(\vec{\omega} \cdot \vec{\omega}
\right) 
\left(\vec{x} \cdot \vec{x} \right) - 
\left(\vec{\omega} \cdot \vec{x} \right)^2 \right],
\label{A}
\nl
g_{0j} & = & - \frac{1}{c} \left(\vec{\omega} \times \vec{x} \right)_j,
\label{B}
\nl
g_{jk} & = & \eta_{jk},
\label{C}
\ee
where $\eta_{\mu \nu}$ represents the Minkowski metric of signature $-
2$.  Equation (\ref{2.1}) has an exact solution to first order in the 
weak-field approximation defined by $g_{\mu \nu}(x) \ = \ \eta_{\mu \nu} +
\gamma_{\mu \nu} (x),$ where the metric deviation $\gamma_{\mu
\nu}$ is a small quantity of first order. 

In fact, a new spinor $\tilde{\psi}(x)$ defined by \cite{14*}
\be
\tilde{\psi}(x) 
\ \equiv \ e^{i \Phi_{\rm S}/\hbar} \, \psi (x),
\label{E}
\ee
where
\be
\Phi_{\rm S} & \equiv & \hbar {\cal P} \int ^x _X dz ^\lambda
\Gamma_\lambda (z),
\label{2.19}
\ee
satisfies the equation 
\be
\left(i \tilde{\gamma}^\mu (x) \nabla_\mu  - 
\frac{mc}{\hbar} \right) \tilde{\psi} (x) \ = \ 0,
\label{2.19a}
\ee
where ${\cal P}$
refers to path ordering and
\be
\tilde{\gamma}^\mu (x) \ \equiv \ 
e^{i \Phi_{\rm S}/ \hbar} 
\gamma^\mu (x)
e^{-i \Phi_{\rm S}/ \hbar}.
\label{2.19b}
\ee
By multiplying (\ref{2.19a}) on the left by $(-i \tilde{\gamma}^\nu (x) \nabla_\nu  - {mc}/{\hbar})$, we obtain the equation \cite{4}
\be
\left(g^{\mu \nu} \nabla_\mu \nabla_\nu + \frac{m^2 c^2}{\hbar^2}\right)
\tilde{\psi}(x) \ = \ 0.
\label{2.20}
\ee 
This last equation can be
solved, in the weak-field approximation, for every component of 
$\tilde{\psi}(x)$ \cite{8}.  Here, $\tilde{\psi}(x)$ is a solution
of a second-order equation and does, of course, contain redundant
solutions.  These can be eliminated by writing the appropriate solution
as
\be
\tilde{\psi}(x) \ = \ 
\left(-i \tilde{\gamma}^\mu (x) \nabla_\mu  - \frac{mc}{\hbar}
\right) 
e^{-i \Phi_{\rm G}/\hbar} \Psi_0,
\label{2.20a}
\ee
where
\be
\Phi_{\rm G} & \equiv & \frac{1}{2} \int_X ^x dz^\lambda
\gamma_{\alpha
\lambda } (z)
P^\alpha
\nn
&   & - \frac{1}{4} \int_X ^x dz^\lambda (\gamma_{\alpha \lambda,
\beta} (z) - \gamma_{\beta \lambda, \alpha} (z) )  L^{\alpha \beta} (z),
\label{2.21}
\ee
and $\Psi_0$ satisfies the Klein-Gordon equation for a free particle in
Minkowski space.

For (\ref{2.21}), 
\be
\left[L^{\alpha \beta} (z), \Psi_0 \right] & \equiv & 
\left\{ (x^\alpha - z^\alpha) P^\beta - (x^\beta - z^\beta) P^\alpha \right\}
\Psi_0,
\label{2.22}
\nl
\left[P^\alpha , \Psi_0 \right] & \equiv & \{ i \hbar \partial^\alpha \}
\Psi_0,
\label{2.23}
\ee
where $P^\mu$ is the momentum operator of the free particle.  In (\ref{2.20a}),
$\Phi_{\rm G}$ plays the role of Berry's phase \cite{15} because
spacetime 
co-ordinates are just parameters and spacetime becomes simply Berry's
parameter space \cite{9}.  The solution of (\ref{2.1}) is, therefore,
$\psi(x) \ = \ \exp \left( - i \Phi_{\rm S}/\hbar \right) 
\tilde{\psi}(x)$ and is exact to first-order in the metric deviation.

It is interesting to notice that, by multiplying (\ref{2.1}) on the left with
$(-i {\gamma}^\nu (x) D_\nu  - {mc}/{\hbar}),$ we obtain the second-order equation
\be
\left(\gamma^\mu(x) \gamma^\nu(x) D_\mu D_\nu + \frac{m^2 c^2}{\hbar^2}\right)
{\psi}(x) \ = \ 0,
\label{ins1}
\ee 
which, on using the relations $[D_\mu, D_\nu] \ = \ -\frac{i}{4} 
\sigma^{\alpha \beta}R_{\alpha \beta \mu \nu}$ and $\sigma^{\mu \nu} \sigma^{\alpha \beta}R_{\mu \nu \alpha \beta} \ = \ 2R$ \cite{14*,14a}, reduces to
\be
\left(g^{\mu \nu} D_\mu D_\nu - \frac{R}{4} + \frac{m^2 c^2}{\hbar^2}\right)
{\psi}(x) \ = \ 0.
\label{ins2}
\ee 
This equation does not contain any spin-curvature coupling for pure gravitational fields ($R$ = 0) and the gyro-gravitational g-factor is, therefore, zero \cite{14b}.
When $R \ = \ - \frac{8\pi G}{c^4}T,$ the g-factor (coefficient of $R$) is $\frac{1}{4}$ \cite{14*}.  An equation that also yields a value for the orbital g-factor can be obtained from (\ref{2.1}) by means of the 
transformation
\be
\psi^\prime (x) \ \equiv \ 
e^{-i \Phi_{\rm G}/\hbar} \,
\psi (x)
.
\label{ins3}
\ee
It is easy to show that $\psi^\prime (x)$ satisfies the equation
\be
\left[i \gamma^\mu (x) \left(D_\mu + \frac{i}{\hbar} (\nabla_\mu \Phi_{\rm G})
\right)
- \frac{mc}{\hbar}\right]\psi^\prime (x) & = & 0.
\label{ins4}
\ee
Equation (\ref{ins4}) can be immediately extended to include the
electromagnetic fields by adding to $\Phi_{\rm G}$ the phase
\be
\Phi_{\rm EM} & \equiv &  \frac{e}{c} \int_X ^x dz^\lambda 
A_\lambda (z),
\label{2.24}
\ee
and can be written in the form
\be
\left[i \gamma^\mu (x) \left(\nabla_\mu + \frac{i}{\hbar} (\nabla_\mu \Theta)
\right)
- \frac{mc}{\hbar}\right]\psi^\prime (x) & = & 0,
\label{ins5}
\ee
where $\Theta \ \equiv \ \Phi_{\rm G} + \Phi_{\rm EM} + \Phi_{\rm S}.$

By operating from the left with $\left[-i \gamma^\nu (x) \left(D_\nu + \frac{i}{\hbar} (\nabla_\nu \Phi_{\rm G})
\right) - {mc}/{\hbar}\right]$ on both sides of (\ref{ins4}), we obtain to first-order in the metric deviation
\be
\left[g^{\mu \nu} D_\mu D_\nu - \frac{i}{2} \sigma^{\mu \nu} \left(
[D_\mu, D_\nu] + i G_{\mu \nu} \right) + \frac{m^2 c^2}{\hbar^2}\right]
{\psi}^\prime (x) \ = \ 0,
\label{ins6}
\ee 
where $\frac{1}{\hbar} (\nabla_\mu \Phi_{\rm G}) \ \equiv \ \kappa_\mu$ and
$G_{\mu \nu} \ \equiv \ \kappa_{\nu;\mu} - \kappa_{\mu;\nu}.$  
On the other hand, $\frac{1}{2} G_{\mu \nu} \ = \ \frac{1}{4} R_{\mu \nu \alpha \beta}L^{\alpha \beta}$ 
and the resulting orbital g-factor is also $\frac{1}{4}.$ This confirms that 
the gyro-gravitational 
ratio of a spin-1/2 particle is 1 as shown in \cite{DE,AU,KA}.

Equation (\ref{ins5}) can now be used to derive a Hamiltonian in general 
co-ordinates, taking the form 
\be
i\hbar c \, \nabla_0 \psi^\prime (x) & = &  (g^{00} (x))^{-1} 
\left[ c \, \gamma^0 (x) \gamma^j(x) \left(-i \hbar \nabla_j\right) + 
mc^2 \gamma^0(x) \right.
\nn
&   &{} \left. + c \, \gamma^0 (x) \gamma^\mu (x) (\nabla_\mu \Theta \right)]\psi^\prime (x) \ = \ H \psi^\prime(x),
\label{2.35*}
\ee
where
\be
g^{00} (x) & = & \left( e^0{}_{ \hat{0}} \right)^2 
\eta^{\hat{0} \hat{0}}
\ = \ \left( 1 + \frac{\vec{a} \cdot \vec{x} }{c^2} \right)^{-2},
\label{2.36*}
\nl
\gamma^0 (x) & = & e^0{}_{ \hat{0}} \gamma^{\hat{0}}
\ = \  \left( 1 + \frac{\vec{a} \cdot \vec{x} }{c^2} \right)^{-1}
 \beta,
\label{2.37*}
\nl
\gamma^0 (x) \gamma^j (x) & = & 
e^0{}_{ \hat{0}} \left( \gamma^{\hat{0}}\gamma^{\hat{\jmath}} 
+ e^j{}_{ \hat{0}} \right)
\nn
& = & \left( 1 + \frac{\vec{a} \cdot \vec{x} }{c^2} \right)^{-2}
\left[ \left( 1 + \frac{\vec{a} \cdot \vec{x} }{c^2} \right)
\alpha^{\hat{\jmath}} - \frac{1}{c}  \epsilon^{jkl} \, \omega_k \, x_l
\right].
\label{2.38*}
\ee
Since $\Phi_{\rm G}$ is correct only to first-order, this is also a constraint on the validity of (\ref{2.35*}).  In what follows, terms of higher order in the metric deviation will be dropped.
Explicit evaluation of $\nabla_\mu \Theta$ shows that
\be
(\nabla_\mu \Theta) & = & \nabla_\mu (\Phi_{\rm EM} + \Phi_{\rm S} +
\Phi_{\rm G})
\ = \  \frac{e}{c} A_\mu  + \hbar \Gamma_\mu + (\nabla_\mu \Phi_{\rm G}),
\label{2.33}
\ee
where
\be
\Gamma_0 & = & - \frac{i}{2c^2} ( \vec{a} \cdot
\vec{\alpha} ) - \frac{1}{2c}  \vec{\omega} \cdot \vec{\sigma}, 
\label{2.32*}
\nl
\Gamma_j & = & 0,
\label{2.33*}
\ee
and
\be
(\nabla_\mu \Phi_{\rm G}) & = &  \frac{1}{2} \gamma_{\alpha \mu} (x)
p^\alpha -
\frac{1}{2} \int_X^x dz^\lambda (\gamma_{\mu \lambda, \beta}(z) - 
\gamma_{\beta \lambda, \mu}(z))p^\beta,
\label{2.34*}
\ee
where $p^\mu$ is the momentum eigenvalue of the free particle.

It follows that, to first-order in $\vec{a}$ and $\vec{\omega}$, the Dirac Hamiltonian in the general co-ordinate frame is
\be
H \ \approx \  c( \vec{\alpha} \cdot \vec{p} ) + mc^2 \beta + V(\vec{x}),
\label{2.34}
\ee
where
\be
V(\vec{x}) & = & \frac{1}{c}( \vec{a} \cdot \vec{x} ) (
\vec{\alpha} \cdot \vec{p} )
+ m(\vec{a} \cdot \vec{x}) \beta - \vec{\omega} \cdot (\vec{L} +
\vec{S}) - \frac{i \hbar}{2c} (\vec{a} \cdot \vec{\alpha})
\nn
&   & {} - e \left( 1 + \frac{\vec{a} \cdot \vec{x} }{c^2} \right)
(\vec{\alpha} \cdot \vec{A}) + \frac{e}{c} \, \vec{\omega} \cdot (\vec{x}
\times \vec{A}) + e \, \varphi 
\nn
&   & {} + c \, \vec{\alpha} \cdot (\vec{\nabla} \Phi_{\rm G}) 
+ c \, (\nabla_0 \Phi_{\rm G}),
\label{2.35}
\ee
the $\vec{\alpha}, \beta, \vec{\sigma}$ matrices are those of
Minkowski space, and $\vec{L} = \vec{x} \times \vec{p}$  and $\vec{S} = {\hbar} \vec{\sigma}/2 $  are the orbital and spin angular momenta, respectively.



\section{Low-Energy Approximation}
\setcounter{equation}{0}

Although the Dirac Hamiltonian as described by 
(\ref{2.34}) and (\ref{2.35})  is useful as is, there is some
benefit in considering approximations which emphasize both the 
low- and high-energy limits in a particle's range of motion.  In this section,
the low-energy approximation is being considered.

According to the FW transformation technique, it is possible to group
the
Dirac Hamiltonian into the form
\be
H  & = & mc^2 \beta + {\cal O} + {\cal E},
\label{3.1}
\ee
where the ``odd'' and ``even'' operators ${\cal O}$ and ${\cal E},$
respectively satisfy $\{ {\cal O}, \beta \} \ = \ {[ {\cal E}, \beta ]} \ = \
0$.  For this derivation, we introduce by hand the anomalous magnetic
moment
\be
\frac{\kappa e \hbar}{2mc} \sigma^{\mu \nu} F_{\mu \nu}
= \frac{\kappa e \hbar}{2mc} ( i \vec{\alpha} \cdot \vec{E}
- \vec{\sigma} \cdot \vec{B} ),
\label{3.2}
\ee
with $\kappa  \equiv (g - 2)/2,$ as another term in $V(\vec{x})$, by
means of the
substitution
\be
mc^2 \beta \ \rightarrow \ \beta\left[mc^2 + 
\frac{\kappa e \hbar}{2mc} ( i \vec{\alpha} \cdot \vec{E}
- \vec{\sigma} \cdot \vec{B} )\right].  
\label{3.2a}
\ee
Then, given
(\ref{2.34}) and (\ref{2.35}), it is possible to identify with (\ref{3.1}),
where
\be
{\cal O} & = & c \, \vec{\alpha} \cdot \left[ 
\left(1 + \frac{ \vec{a} \cdot \vec{x}}{c^2} \right) 
\vec{\pi} + (\vec{\nabla} \Phi_{\rm G}) - \frac{i \kappa e
\hbar}{2mc^2} \beta \vec{E} - \frac{i\hbar}{2c^2} \vec{a} \right],
\label{3.3}
\nl
{\cal E} & = & \left[ m (\vec{a} \cdot \vec{x})  - 
\frac{ \kappa e \hbar}{2mc}\left(1 + \frac{ \vec{a} \cdot
\vec{x}}{c^2} \right)  (\vec{\sigma} \cdot \vec{B}) \right] \beta
\nn
&   & {}  - \vec{\omega} \cdot (\vec{x} \times \vec{\pi}) -
\vec{\omega}
\cdot \left(\frac{\hbar}{2} \vec{\sigma} \right) + e\varphi 
+ c (\nabla_0 \Phi_{\rm G}),
\label{3.4}
\ee
where $\vec{\pi} \ = \ \vec{p} - e \vec{A}/c.$

Following the procedure given by Bjorken and Drell \cite{13}, the
transformed
Hamiltonian is represented by a series expansion of $S,$ according to a
unitary transformation
\be
H^\prime & = & UHU^{-1}
\nn
& \approx & H + i [S, H] - \frac{1}{2} \left[S, [S, H] \right] -
\frac{i}{6} \left[S, \left[S, [S, H] \right] \right]
\nn
&   & {} + \frac{mc^2}{24} \left[ S,  \left[S, \left[S, [S, \beta] \right]
\right]       \right] - \hbar \dot{S} - \frac{i\hbar}{2} [S, \dot{S} ] ,
\label{3.5}
\ee
and $S \ = \ {\rm O}(1/m)$ is the Hermitian exponent of a unitary
transformation operator $U \ \equiv \ \exp(iS).$  By three successive
applications of (\ref{3.5}) for the choice 
\be
S \ \equiv \ S_{\rm FW} \ = \  - \frac{i}{2mc^2} \beta  {\cal O},
\label{3.6}
\ee
the final transformed Hamiltonian becomes
\be
H_{\rm FW} & = & mc^2 \beta + {\cal E}^\prime
\nn
& = & \beta \left( mc^2 + \frac{1}{2mc^2} {\cal O}^2 -
\frac{1}{8m^3 c^6} {\cal O}^4 \right) + {\cal E}
\nn
&   & {} - \frac{1}{8m^2 c^4} \left[ {\cal O}, [ {\cal O},
{\cal E} ] \right] - \frac{i \hbar}{8m^2 c^4} [ {\cal O},
\dot{\cal O} ].
\label{3.7}
\ee

To determine the gravitational corrections in (\ref{3.7}), it is necessary to isolate the external
electromagnetic potentials within the definition of odd and even
operators.  This implies that ${\cal O} \ \equiv \  {\cal O}_0 + {\cal
O}_1$ and ${\cal E} \ \equiv \ {\cal E}_0 + {\cal E}_1,$ where 
\be
{\cal O}_0 & = & c \, \vec{\alpha} \cdot \vec{\pi},
\label{3.8}
\nl
{\cal E}_0 & = & e \varphi.
\label{3.9}
\ee
Therefore,
\be
{\cal O}_1 & = & c \, \vec{\alpha} \cdot \left[ 
\left(\frac{ \vec{a} \cdot \vec{x}}{c^2} \right) \vec{\pi}
 + (\vec{\nabla} \Phi_{\rm G}) - \frac{i \kappa e
\hbar}{2mc^2} \beta \vec{E} - \frac{i\hbar}{2c^2} \vec{a} \right],
\label{3.10}
\nl
{\cal E}_1 & = & \left[ m (\vec{a} \cdot \vec{x}) - 
\frac{ \kappa e \hbar}{2mc}\left(1 + \frac{ \vec{a} \cdot
\vec{x}}{c^2} \right)  (\vec{\sigma} \cdot \vec{B}) \right] \beta
\nn
&   & {}  - \vec{\omega} \cdot (\vec{x} \times \vec{\pi}) -
\vec{\omega}
\cdot \left(\frac{\hbar}{2} \vec{\sigma} \right) + c (\nabla_0 \Phi_{\rm G}).
\label{3.11}
\ee

Neglecting the ${\cal O}^4$ contribution and considering only terms up
to first-order in $\vec{a}, \ \vec{\omega},$ and $1/m^2,$ it follows that
\be
{\cal O}^2 & = & {\cal O}^2_0 + {\cal O}^2_1 + \{{\cal O}_0 , {\cal
O}_1 \} ,
\label{3.12}
\nl
\left[{\cal O} , [ {\cal O}, {\cal E} ] \right] & \approx
& \left[{\cal O}_0 , [ {\cal O}_0, {\cal E}_0 ] \right]
+ \left[{\cal O}_0 , [ {\cal O}_1, {\cal E}_0 ] \right]
+ \left[{\cal O}_0 , [ {\cal O}_0, {\cal E}_1 ] \right]
+ \left[{\cal O}_1 , [ {\cal O}_0, {\cal E}_0 ] \right],
\label{3.13}
\nl
{ [ {\cal O}, \dot{\cal O} ]} & \approx &  
[ {\cal O}_0, \dot{\cal O}_0 ] 
+ [ {\cal O}_0, \dot{\cal O}_1 ]
+ [ {\cal O}_1, \dot{\cal O}_0 ].
\label{3.14}
\ee

>From the zeroth-order terms in (\ref{3.7}), it can be shown that \cite{13}
\be
H_{\rm FW(0)} & = & mc^2 \beta + \frac{1}{2mc^2} \beta \, {\cal
O}_0^2 
+e \varphi - \frac{1}{8m^2 c^4} \left[ {\cal O}_0, [ {\cal 
O}_0 , {\cal E}_0 ] \right] - \frac{i \hbar}{8m^2 c^4} [ {\cal 
O}_0 , \dot{\cal O}_0 ]
\nn
& = &  mc^2 \beta + \left[ \frac{1}{2m} \pi^2 -
\frac{e \hbar}{2mc} \vec{\sigma} \cdot \vec{B} \right] \beta - \frac{e
\hbar}{4m^2 c^2}
\vec{\sigma} \cdot (\vec{E} \times \vec{\pi})
\nn
&  & {} - \frac{e\hbar^2}{8m^2 c^2} \left[(\vec{\nabla} \cdot \vec{E}) +
i \vec{\sigma} \cdot (\vec{\nabla} \times \vec{E}) \right] 
+ e \varphi,
\label{3.15}
\ee
where the third term coupled to $\beta$ is the magnetic
dipole energy, and the following term is the spin-orbit energy.  

Neglecting the time-dependent contributions from (\ref{3.14}) and considering only those terms up to second-order in $\vec{\pi}$, it follows that
\be
{\cal O}_1^2 & = & - \frac{i \kappa e \hbar}{m} \, \beta \left[
\left(\frac{\vec{a} \cdot \vec{x}}{c^2}\right) (\vec{E} \cdot \vec{\pi})
+ \vec{E} \cdot (\vec{\nabla} \Phi_{\rm G}) \right] -
\frac{\kappa e \hbar^2}{2mc^2} \, \beta (\vec{a} \cdot \vec{E})
\nn
&   &{} - \frac{\kappa e \hbar^2}{2m} \left(\frac{\vec{a} \cdot \vec{x}}{c^2}\right) \beta (\vec{\nabla} \cdot \vec{E}) -
\frac{i \kappa e \hbar^2}{2m} \left(\frac{\vec{a} \cdot \vec{x}}{c^2}\right) 
\beta \, \vec{\sigma} \cdot (\vec{\nabla} \times \vec{E}), 
\label{3.16}
\ee
\be
\{{\cal O}_0 , {\cal O}_1 \} & = & (\vec{a} \cdot \vec{x}) \pi^2 +
2c^2 ((\vec{\nabla} \Phi_{\rm G}) \cdot \vec{\pi}) -
\frac{i \kappa e \hbar}{m} \, \beta (\vec{E} \cdot \vec{\pi}) - 
i \hbar (\vec{a} \cdot \vec{\pi})
\nn
&  & {}+ c^2 \vec{\pi} \left(\frac{\vec{a} \cdot \vec{x}}{c^2}\right)
\cdot \vec{\pi} - i \hbar c^2 (\nabla^2 \Phi_{\rm G}) - 
\frac{\kappa e \hbar^2}{2m} \, \beta (\vec{\nabla} \cdot \vec{E})
- \frac{\kappa e \hbar}{m} \beta \, \vec{\sigma} \cdot (\vec{E} \times
\vec{\pi})
\nn
&  & {} + \hbar \, \vec{\sigma} \cdot (\vec{a} \times \vec{\pi}) -
2e \hbar c \left(\frac{\vec{a} \cdot \vec{x}}{c^2}\right)
\vec{\sigma} \cdot (\vec{\nabla} \times \vec{A}) -
\frac{i \kappa e \hbar^2}{2m} \, \beta \, \vec{\sigma} \cdot (\vec{\nabla}
\times \vec{E}),
\label{3.17}
\ee

\be
\left[{\cal O}_0 , [ {\cal O}_1, {\cal E}_0 ] \right] & = &
i e \hbar^2 \, \vec{a} \cdot \vec{\nabla}\varphi - e \hbar^2 (\vec{a} \cdot \vec{x}) \nabla^2 \varphi - 2 e \hbar (\vec{a} \cdot \vec{x}) 
\vec{\sigma} \cdot (\vec{\nabla}\varphi \times \vec{\pi}), 
\label{3.18}
\ee

\be
\left[{\cal O}_0 , [ {\cal O}_0, {\cal E}_1 ] \right] & = &
-4i mc^2 \hbar \, \beta (\vec{a} \cdot \vec{\pi}) - \hbar^2 c^3
\nabla^2(\nabla_0 \Phi_{\rm G}) - i e \hbar^2 c \, \vec{\sigma} \cdot
(\vec{\omega} \times (\vec{\nabla} \times \vec{A}))
\nn
&  &{}+ 2 mc^2 \hbar \, \beta \, \vec{\sigma} \cdot (\vec{a} \times \vec{\pi})
+ 4mc^2 \, \beta (\vec{a} \cdot \vec{x}) \pi^2 - 4emc \hbar \, \beta
(\vec{a} \cdot \vec{x}) \vec{\sigma} \cdot (\vec{\nabla} \times \vec{A})
\nn
&  &{} - 2 \hbar c^3 \, \vec{\sigma} \cdot (\vec{\nabla}(\nabla_0 \Phi_{\rm G})
\times \vec{\pi}) + 2 \hbar c^2 \, \vec{\sigma} \cdot (\vec{\omega} \times
\vec{\pi}) \times \vec{\pi},
\label{3.19}
\ee

\be
\left[{\cal O}_1 , [ {\cal O}_0, {\cal E}_0 ] \right] & = & 
- \hbar^2 ec^2 \left(\frac{\vec{a} \cdot \vec{x}}{c^2}\right) \nabla^2 \varphi
- 2 \hbar ec^2 \left(\frac{\vec{a} \cdot \vec{x}}{c^2}\right) \vec{\sigma}
\cdot (\vec{\nabla}\varphi \times \vec{\pi}) 
\nn
&  &{} + 2 \hbar ec^2
\vec{\sigma} \cdot ((\vec{\nabla}\Phi_{\rm G}) \times \vec{\nabla}\varphi) 
- i \hbar^2 e \, \vec{\sigma} \cdot (\vec{a} \times \vec{\nabla}\varphi).
\label{3.20}
\ee

After neglecting the non-Hermitian terms in (\ref{3.16}) - (\ref{3.20}), it becomes evident that the low-energy approximation for the Dirac Hamiltonian is

\be
H_{\rm FW} & \approx & 
mc^2 \beta + \left[ \frac{1}{2m} \pi^2 -
\frac{e \hbar}{2mc} \vec{\sigma} \cdot \vec{B} \right] \beta - \frac{(g-1)e
\hbar}{4m^2 c^2}
\vec{\sigma} \cdot (\vec{E} \times \vec{\pi})
\nn
&  & {} - \frac{e\hbar^2}{8m^2 c^2} \left[(\vec{\nabla} \cdot \vec{E}) +
i \vec{\sigma} \cdot (\vec{\nabla} \times \vec{E}) \right] 
+ e \varphi
\nn
&  &{} +
\left[ m (\vec{a} \cdot \vec{x}) - 
\frac{e \hbar}{2mc}\left\{ \kappa \left(1 + \frac{ \vec{a} \cdot
\vec{x}}{c^2} \right) + \left( \frac{ \vec{a} \cdot \vec{x}}{c^2} \right) 
\right\} 
(\vec{\sigma} \cdot \vec{B}) \right] \beta
\nn
&   & {}  - \vec{\omega} \cdot (\vec{x} \times \vec{\pi}) -
\vec{\omega}
\cdot \left(\frac{\hbar}{2} \vec{\sigma} \right) + c (\nabla_0 \Phi_{\rm G})
\nn
&  & + \frac{1}{2m} \, \beta \vec{\pi} \left(\frac{\vec{a} \cdot \vec{x}}{c^2}\right) \cdot \vec{\pi} + \frac{\hbar}{4mc^2} \, \beta \,
\vec{\sigma} \cdot (\vec{a} \times \vec{\pi}) - \frac{\hbar}{4m^2c^2} 
\vec{\sigma} \cdot (\vec{\omega} \times \vec{\pi}) \times \vec{\pi}
\nn
&  &{} - \frac{\kappa e \hbar^2}{4m^2 c^2} \left(1 + \frac{\vec{a} \cdot \vec{x}}{c^2}\right) \left[(\vec{\nabla} \cdot \vec{E}) + i \, \vec{\sigma}
\cdot (\vec{\nabla} \times \vec{E}) \right]
\nn
&  &{} - \frac{\kappa e \hbar^2}{4m^2 c^4} (\vec{a} \cdot \vec{E})
+ \frac{e \hbar^2}{4m^2 c^2} \left(\frac{\vec{a} \cdot \vec{x}}{c^2}\right) 
\nabla^2 \varphi 
\nn
&  & {} + \frac{1}{m} \, \beta \, (\vec{\nabla} \Phi_{\rm G}) \cdot \vec{\pi} +
\frac{e \hbar}{4m^2 c^2} \, \vec{\sigma} \cdot \left(
\vec{\nabla}\varphi \times \left[(\vec{\nabla} \Phi_{\rm G}) + 
\left(\frac{\vec{a} \cdot \vec{x}}{c^2}\right) \vec{\pi} \right] \right)
\nn
&  & {} + \frac{\hbar^2}{8m^2 c} \nabla^2(\nabla_0 \Phi_{\rm G}) +
\frac{\hbar}{4m^2 c} \, \vec{\sigma} \cdot (\vec{\nabla}(\nabla_0 \Phi_{\rm G})
\times \vec{\pi}). 
\label{3.21}
\ee
The occurrence of non-Hermitian terms, here neglected, is a well known 
phenomenon,
likely connected with the breakdown of the single-particle interpretation of the 
Dirac equation in the presence of time-dependent inertial and gravitational 
fields \cite{!,HU}.

%


\section{High-Energy Approximation}
\setcounter{equation}{0}

Though used less often than the FW-transformation, the
CT-transformation follows the same mathematical principles as the
former to arrive at the high-energy limit for the Dirac Hamiltonian. 
Although the FW-transformation can be successfully applied when 
non-trivial potential energy terms are present, it is far from obvious that
the CT-transformation can do the same.  This is because it is not clear
how to classify the Dirac Hamiltonian into a high-energy analogue of
odd
and even operators, as found in the FW approach, in order to
systematically remove the undesired terms.  An unfortunate consequence
of this impasse is that it becomes impossible to analyze the motion and
properties of fast-moving massive particles in the presence of fields
without arbitrarily setting its mass equal to zero within the Hamiltonian. 
Clearly, this precludes any opportunity to compare the behaviour of
these particles with that of strictly massless particles. 

It is shown below, however, that it is possible to derive a high-energy
approximation of the generalized Dirac Hamiltonian for a spin-1/2
particle moving in a potential defined by both electromagnetic
and gravitational fields.  For this derivation, the final expression
for the Hamiltonian is also in terms of a general co-ordinate frame. The
steps now proceed backward from the Dirac equation or corresponding
second-order equation in locally Minkowski space.

Given that
\be
i \hbar c \, \nabla_{\hat{0}} \Psi_0 & = & H_0 \Psi_0
\nn
& = & \left[- c \,
\alpha^{\hat{\jmath}} P_{\hat{\jmath}} + mc^2 \beta \right] \Psi_0,
\label{4.1}
\ee
it is possible to define a new wavefunction $\Psi^\prime_0 \ \equiv \ \exp (i
S_{\rm CT}) \Psi_0,$ where
\be
S_{\rm CT} & \approx & - \frac{imc}{2p^2} \beta
(\alpha^{\hat{\jmath}} P_{\hat{\jmath}}), 
\label{4.2}
\ee
and $[ \nabla_{\hat{\mu}}, S_{\rm CT} ] \ = \ 0$.  It then follows from
the CT-transformation that
\be
i \hbar c \, \nabla_{\hat{0}} \Psi ^\prime_0 & = & \left[ e^{iS_{\rm
CT}}
H_0 \, e^{-iS_{\rm CT}} \right] \Psi ^\prime_0
\nn
& \approx & H_{\rm CT(0)} \Psi ^\prime_0,
\label{4.3}
\ee
where 
\be
H_{\rm CT(0)} & \approx & - \frac{E}{p} (\alpha^{\hat{\jmath}} P_{\hat{\jmath}})
\label{4.4}
\ee
is the CT Hamiltonian for the free particle.  Since $\Psi_0$ can be
related to $\tilde{\psi}(x)$ by (\ref{2.20a}), it can be shown that
\be
\Psi^\prime_0 & = & e^{iS_{\rm CT}} \Psi_0
\nn
& = & 
e^{iS_{\rm CT}} \left[e^{i \Theta / \hbar} \psi^{\prime
\prime} (x)
\right],
\label{4.5}
\ee
where
\be
\psi^{\prime \prime} (x) & = & e^{-i\Phi_{\rm S}/\hbar}
\left(-i\gamma^\mu (x) \nabla_\mu - \frac{mc}{\hbar} \right)^{-1}
e^{i\Phi_{\rm S}/\hbar} \, \psi (x).
\label{4.5a}
\ee
Substituting (\ref{4.5}) into (\ref{4.3}), we obtain
\be
- i \hbar c \left[ \frac{E}{pc} \left( e^{-iS_{\rm CT}} \, 
\alpha^{\hat{\jmath}} \, e^{iS_{\rm CT}}  \right) \nabla_{\hat{\jmath}}
+
\nabla_{\hat{0}} \right] e^{i {\Theta}/{\hbar}} 
\psi^{\prime \prime} (x) & = & 0.
\label{4.6}
\ee
By using the vierbeins, it becomes possible to describe (\ref{4.6}) in
terms of the general co-ordinate frame, so that 
\be
- i \hbar c \left[ \frac{E}{pc} \left( e^{-iS_{\rm CT}} \, 
\alpha^{\hat{\jmath}} \, e^{iS_{\rm CT}} \right)
e^\mu{}_{\hat{\jmath}} +
e^\mu{}_{\hat{0}} \right] \nabla_\mu \left(e^{i {\Theta}/\hbar} 
\psi^{\prime \prime} (x) 
\right) & = & 0.
\label{4.6a}
\ee

It is a straightforward matter to evaluate the transformation of
$\alpha^{\hat{\jmath}}.$  Though it is possible to perform the expansion
for
higher-order terms, it is only necessary
to consider the zeroth- and first-order terms.  It follows that
\be
e^{-iS_{\rm CT}} \, \alpha^{\hat{\jmath}} \, e^{iS_{\rm CT}}  &
\approx &
\alpha^{\hat{\jmath}} - i \, [S_{\rm CT} , \alpha^{\hat{\jmath}}]
\nn
& \approx & \alpha^{\hat{\jmath}} + \left( \frac{mc}{p^2} \right) \beta
P^{\hat{\jmath}}.
\label{4.7}
\ee
Substituting (\ref{4.7}) into (\ref{4.6}), we obtain the
expression
\be
i \hbar c \, \nabla_0 \psi^{\prime \prime} (x) & = & 
\frac{E}{p} \left(1 + \frac{\vec{a} \cdot \vec{x}}{c^2} \right)
\left[\vec{\alpha}^{\prime \prime}  \cdot \vec{p} + 
\vec{\alpha}^{\prime \prime}  \cdot \vec{\nabla} \Theta
+ \left(\frac{mc}{p^2} \right) \beta^{\prime \prime}  \left[ (\vec{p} \cdot \vec{p}) 
+ \vec{p} \cdot \vec{\nabla} \Theta \right] \right] \psi^{\prime \prime}(x)
\nn
&   & {} - \vec{\omega} \cdot \left[\vec{x} \times \left(\vec{p} +
\vec{\nabla} \
\Theta \right) \right] \psi^{\prime \prime}(x) 
+ c \, \nabla_0 \Theta \, \psi^{\prime \prime}  (x)
\nn
& = & {} H_{\rm CT} \, \psi^{\prime \prime} (x),
\label{4.8}
\ee
where
\be
\vec{\alpha}^{\prime \prime} & = & e^{-i \Phi_{\rm S}/\hbar} \, 
\vec{\alpha} \,  
e^{i \Phi_{\rm S}/\hbar},
\label{4.8a}
\nl
\beta^{\prime \prime} & = & e^{-i \Phi_{\rm S}/\hbar} \, \beta \,  
e^{i \Phi_{\rm S}/\hbar}.
\label{4.8b}
\ee
To first-order in $\vec{a}$ and $\vec{\omega}$, the
$\nabla_\mu \Theta$ contribution is given by (\ref{2.33}).  
The only remaining term to be evaluated is $\vec{p} \cdot
\vec{\nabla} 
\Theta.$  It follows that
\be
\vec{p} \cdot \vec{\nabla} \Theta & = & - \frac{e}{c} 
\left(\vec{p} \cdot \vec{A} \right) + \vec{p} \cdot (\vec{\nabla} \Phi_{\rm G}).
\label{4.9}
\ee
Therefore, 
the final expression for the CT-Hamiltonian with electromagnetic
and gravitational fields present is
\be
H_{\rm CT} & \approx & \frac{E}{p} (\vec{\alpha}^{\prime \prime} \cdot \vec{p}) + V(\vec{x})_{\rm CT},
\label{4.10}
\ee
where
\be
V(\vec{x})_{\rm CT} & \approx & \frac{E}{pc^2} ( \vec{a} \cdot
\vec{x}
) (\vec{\alpha}^{\prime \prime}  \cdot \vec{p} ) +
\frac{E}{p} \left(1 + \frac{\vec{a} \cdot \vec{x}}{c^2} \right) \left(
\frac{mc}{p^2} \right) \beta^{\prime \prime}  (\vec{p} \cdot \vec{p})
  - \vec{\omega} \cdot (\vec{L} + \vec{S}) - \frac{i\hbar}{2c} (\vec{a} 
\cdot \vec{\alpha})
\nn
&   & {} - \frac{Ee}{pc} \left(1 + \frac{\vec{a} \cdot \vec{x}}{c^2}
\right)
\vec{\alpha}^{\prime \prime}  \cdot \vec{A} + \frac{e}{c} \,
\vec{\omega} \cdot (\vec{x} \times \vec{A}) + e \, \varphi 
+ \frac{E}{p} \, \vec{\alpha}^{\prime \prime} \cdot (\vec{\nabla} \Phi_{\rm G})
+ c \, (\nabla_0 \Phi_{\rm G})
\nn
&   & {} + \frac{E}{p} \left(1 + \frac{\vec{a}\cdot \vec{x}}{c^2} \right) 
\left(\frac{mc}{p^2} \right) \beta^{\prime \prime}  \left[\vec{p} \cdot (\vec{\nabla} 
\Phi_{\rm G}) - \frac{e}{c} 
\left(\vec{p} \cdot \vec{A}\right) \right] 
.
\label{4.11}
\ee
>From (\ref{2.19}), (\ref{2.32*}) - (\ref{2.33*}), (\ref{4.8a}) and (\ref{4.8b}), the first-order expansions of $\vec{\alpha}^{\prime \prime}$ and $\beta^{\prime \prime}$ are
\be
\vec{\alpha}^{\prime \prime} & \approx & \vec{\alpha} + {\cal P}
\int_X^x \, dz^0 \left[\frac{i}{c^2} (\vec{a} \times \vec{\sigma}) +
\frac{1}{2c} (\vec{\omega} \times \vec{\alpha})\right],
\label{4.12}
\nl
\beta^{\prime \prime} & \approx & \beta + \frac{1}{c^2} \, {\cal P}
\int_X^x \, dz^0 \, \beta (\vec{a} \cdot \vec{\alpha}).
\label{4.13}
\ee
 

\section{Conclusions}
\setcounter{equation}{0}

The study of inertia/gravity requires in general knowledge of the 
phase factors $\Phi_{\rm G}$, along with $\Phi_{\rm S}$
and
the Hamiltonian.  The procedure followed gives both.  
It also leads to a solution of the Dirac equation that, as $\Phi_{\rm G}$, 
is exact to first-order in the metric deviation.  As well, $\Phi_{\rm S}$
is exact. 

The Hamiltonian for both the non-relativistic and extreme relativistic
cases are represented by (\ref{2.34}) -- (\ref{2.35}) 
and (\ref{4.10}) -- (\ref{4.11}), respectively.  Low- and high-energy 
approximations can be further developed by following well-known
procedures.
The low-energy approximation only has been derived in detail.  A comparison of 
(\ref{2.35}) with (\ref{4.11}) indicates differences only in terms
proportional
to the mass.  This is understandable in view of the different expansions
of
the energy implied by the corresponding approximations.

Some of the terms that appear in (\ref{2.35}) and (\ref{4.11}) are
identical
and can be identified with corresponding terms that appear in the
Hamiltonian
of Hehl and Ni.  One such
instance
is represented by the term $-\vec{\omega} \cdot \vec{S}$, the
Mashhoon effect \cite{**,3*},
wrongly interpreted by Bell and Leinaas as a version of the Unruh effect.
The term $-\vec{\omega} \cdot \vec{L}$ is the Page-Werner effect
\cite{4*}, while 
$m(\vec{a} \cdot \vec{x}) \beta$ represents the Bonse-Wroblewski
effect \cite{5*}.  Both
effects have been tested experimentally.  Also present is the term
$(\vec{a} 
\cdot \vec{x})(\vec{\alpha} \cdot \vec{p})/c$, which is an
energy-momentum
redshift. While several terms in (\ref{4.11}) vanish in the
limit $m \ \rightarrow \ 0$, the Mashhoon effect is not affected by this
limit.  This leads one to conclude that the rotation-helicity effects
discussed by Cai and Papini \cite{4,5} for massive neutrinos persist in the
vanishing mass limit.  The Mashhoon effect is obviously a prime
candidate for experiments with accelerators and interferometers.
In the latter case, $\Theta$ can be applied to a spacetime loop that is
effectively closed because of the coherence of the particle
wavefunctions.
The result is manifestly gauge invariant and is given by
\be
\Theta \ = \ \frac{1}{4} \int_\Sigma R_{\mu \nu \alpha \beta} \, 
J^{\alpha \beta} \, d \tau^{\mu \nu} + \frac{e}{c} \int_\Sigma F_{\mu \nu} d\tau^{\mu \nu},
\label{5.1}
\ee
where $J^{\alpha \beta}$ is the total angular momentum of the particle, 
$R_{\mu \nu \alpha \beta}$ is the linearized Riemann tensor, $F_{\mu \nu}$ is 
the electromagnetic field tensor and
$\Sigma$ is a surface bounded by the loop. Contributions by the second order derivatives of the metric 
therefore appear in measurable phases also in the case of inertial fields. Eq.(\ref{5.1}) 
clearly indicates that the phase shifts of quantum interferometry depend on the masses of the
particles involved and that a strong form of the equivalence principle cannot be 
present at the quantum level \cite{LA}. Nonetheless Eq.(\ref{2.20}), on which our solution 
is based, still implies that gravitational fields can be simulated locally by acceleration fields \cite{MA}.

Eq.(\ref{3.21}) for the low-energy Hamiltonian can be now compared with the 
results obtained by other authors. Let us neglect, for simplicity, the anomalous 
magnetic moment contributions introduced in Section 3. The Bonse-Wroblewski, Page-Werner and Mashhoon 
terms can be immediately 
recognized by inspection. They correspond to the eighth, tenth and 
eleventh terms respectively. The fifteenth term contains electromagnetic and momentum 
corrections to the Mashhoon effect. The thirteenth term represents 
the redshift effect of the kinetic energy already mentioned, but here in the company  
of its electromagnetic corrections. These also appear in the new inertial spin-orbit 
term found by Hehl and Ni (the fourteenth term). The fourth and sixth terms 
represent spin-orbit coupling and are discussed, for instance, by Bjorken and Drell.
The third term is also wellknown and represents the magnetic dipole interaction. 
The Darwin term is the fifth and the nineteenth represents an acceleration 
correction to it. All remaining terms are proportional to the derivatives of $\Phi_{\rm G}$ 
(see Eq.(\ref{2.34*}). Among these $c (\nabla_0 \Phi_{\rm G}) + \frac{1}{m} \, \beta \, (\vec{\nabla} \Phi_{\rm G}) \cdot \vec{\pi}$ 
appear to dominate. The integral-dependent part of (\ref{2.34*}) yields contributions 
that are small for small paths and low particle momenta. They produce, in general, 
curvature contributions for closed spacetime paths, as in interferometry, and lead to 
(\ref{5.1}) above. The largest contributions come from the first part of (\ref{2.34*}) 
which contains the terms $\frac{1}{2} m c^2 \gamma_{00} (x)$ and $m c \gamma_{0i} (x)
p^i$ already discussed by De Witt and Papini in connection with the 
behaviour of superconductors in weak inertial and gravitational fields.

In view of the above, Eq.(\ref{ins5}) appears remarkably successful in dealing with
all the inertial and gravitational effects discussed in the literature.





\begin{thebibliography}{99}
\bibitem{***} Audretsch J, Hehl F W and L\"{a}mmerzahl C 1992 {\it
Relativistic Gravity Research with Emphasis on Experiments and
Observations--Proc. of the Bad Honnef School on Gravitation} ed J
Ehlers and G Sch\"{a}fer (Springer-Verlag:  New York)
\bibitem{1} Hildebrandt A F, Saffren M M 1965 {\it Proc. 9th Int. Conf.
on Low-Temp. Phys. Pt. A} 459 \\
Hendricks J B and Rorschach H E Jr {\it ibid.} 466 \\
Bol M and Fairbank W M {\it ibid.} 471 \\
Zimmerman J E and Mercereau J E 1965 {\it Phys. Rev. Lett.} {\bf 14}
887
\bibitem{3}Colella R, Overhauser A W and Werner S A 1975 {\it Phys.
Rev. Lett} {\bf 34} 1472 \\
Werner S A, Staudenman J-L and Colella R 1979 {\it Phys. Rev. Lett.}
{\bf 42} 1103 \\
Bonse V and Wroblewski T 1983 {\it Phys. Rev. Lett.} {\bf 51} 1401
\bibitem{DW} DeWitt B S 1966 {\it Phys. Rev. Lett.} {\bf 16} 1092
\bibitem{2} Papini G 1966 {\it Nuovo Cimento} {\bf 45} 66 \\ 1966 {\it Phys. Lett.} 
{\bf 23} 418 \\ 1967 {\it Nuovo Cimento} {\bf 52} 136
\bibitem{LA} L\"{a}mmerzahl C 1996 {\it General Relativity and Gravitation} {\bf 28} 
1043
\bibitem{MA} Mannheim Ph D {\it gr-qc/9810087}
\bibitem{DE} De Oliveira C G and Tiomno J 1962 {\it Nuovo Cimento} 
{\bf 24} 672 
\bibitem{14a} Peres A 1962 {\it Suppl. Nuovo Cimento} {\bf 24} 389
\bibitem{4} Cai Y Q and Papini G 1991 {\it Phys. Rev. Lett.} {\bf 66}
1259; 1992 {\it Phys. Rev. Lett.} {\bf 68} 3811
\bibitem{5} Papini G 1994 {\it Proc. of the 5th Canadian Conference on
General Relativity and Relativistic Astrophysics} ed R B Mann and R G
McLenaghan (World Scientific:  Singapore) 107
\bibitem{6} Opat G I 1983 {\it Proc. 3rd M. Grossmann Meeting in
General Relativity} ed H Ning (Science Press, North-Holland: 
Amsterdam)
\bibitem{7} Werner S A and Kaiser H 1990 {\it Quantum Mechanics in
Curved Space-Time} ed J Audretsch and V de Sabbata (Plenum Press: 
New York) 1
\bibitem{8} Cai Y Q and Papini G 1989 {\it Class. Quantum Grav.} {\bf
6} 407
\bibitem{9} Cai Y Q and Papini G 1989 {\it Mod. Phys. Lett.} {\bf A4}
1143; 1990 {\it Class. Quantum Grav.} {\bf 7} 269; 1990 {\it Gen. Rel.
Grav.} {\bf 22} 259 \\
Papini G 1990 {\it Quantum Mechanics in Curved Space-Time} ed J
Audretsch and V de Sabbata (Plenum Press:  New York) 473
\bibitem{10} Cai Y Q, Lloyd and Papini G 1993 {\it Phys. Lett.}
{\bf A178} 225
\bibitem{11} Bell J S and Leinaas J 1987 {\it Nucl. Phys.} {\bf B284}
488
\bibitem{!} Parker J L 1980 {\it Phys. Rev.} {\bf D22} 1922
\bibitem{12} Hehl F W and Ni W-T 1990 {\it Phys. Rev.} {\bf D42} 2045
\bibitem{CT} Cini M and Touschek B 1958 {\it Nuovo Cimento} {\bf
7} 422
\bibitem{13} Bjorken J D and Drell S D 1964 {\it Relativistic Quantum
Mechanics} (McGraw-Hill:  San Francisco)
\bibitem{14} De Felice F and Clarke C J S 1990 {\it Relativity on
Curved Manifolds} (Cambridge University Press:  New York)
\bibitem{14*} Pagels H 1965 {\it Ann. Phys. (N.Y.)} {\bf 31} 64
\bibitem{14b} Peres A 1963 {\it Nuovo Cimento} {\bf 28} 1091
\bibitem{15} Berry M V 1984 {\it Proc. of the Royal Soc., London} {\bf
A392} 45
\bibitem{AU} Audretsch J 1981 {\it J. Phys. A: Math. Gen.} {\bf 14} 411 
\bibitem{KA} Kannenberg L 1977 {\it Ann. Phys. (N.Y.)} {\bf 103} 64
\bibitem{HU} Huang J C  1994 {\it Ann. Physik} {\bf 3} 53
\bibitem{**} Mashhoon B 1988 {\it Phys. Rev. Lett.} {\bf 61} 2639
\bibitem{3*} Mashhoon B 1989 {\it Phys. Lett.} {\bf A139} 103
\bibitem{4*} Werner S A, Staudenmann, J-L and Colella R 1979 {\it Phys.
Rev. Lett.} {\bf 42} 1103 \\
Page L A 1975 {\it ibid.} {\bf 35} 543
\bibitem{5*} Bonse V and Wroblewski T 1983 {\it Phys. Rev. Lett.}
{\bf 51} 1401
\end{thebibliography}
\end{document}